# Roman CCS White Paper

# Identifying high-redshift pair-instability supernovae by adding sparse F213 filter observations

**Roman Core Community Survey:** *High Latitude Time Domain Survey*

**Scientific Categories:** *stellar physics and stellar types*

**Additional scientific keywords:** *supernovae*

**Submitting Author:**
Name: Takashi Moriya
Affiliation: National Astronomical Observatory of Japan
Email: takashi.moriya@nao.ac.jp

**List of contributing authors** (including affiliation and email):
Ori D. Fox (STScI, ofox@stsci.edu), Robert Quimby (SDSU, rquimby@sdsu.edu), Steve Schulze (OKC, steve.schulze@fysik.su.se), Ashley Villar (PSU, vav5084@psu.edu), Armin Rest (STScI, arest@stsci.edu), Norman Grogin (STScI, nagrogin@stsci.edu), Sebastian Gomez (STScI, sgomez@stsci.edu), David Rubin (UH, drubin@hawaii.edu), Matt Siebert (STScI, msiebert@stsci.edu), Susan Kassin (STScI, kassin@stsci.edu), Eniko Regos (Konkoly, regos@konkoly.hu), Lou Strolger (STScI, strolger@stsci.edu), Anton Koekemoer (STScI, koekemoer@stsci.edu), Steven Finkelstein (UT Austin, stevenf@astro.as.utexas.edu), Suvi Gezari (STScI, sgezari@stsci.edu), Seppo Mattila (Univ. of Turku, sepmat@utu.fi), Tea Temim (Princeton, temim@astro.princeton.edu), Melissa Shahbandeh (STScI, mshahbandeh@stsci.edu), Bob Williams (STScI, wms@stsci.edu), Ting-Wan Chen (TUM, janet.chen@astro.su.se), Isobel Hook (Lancaster University, i.hook@lancaster.ac.uk), Justin Pierel (STScI, jpierel@stsci.edu), Masami Ouchi (NAOJ/UTokyo, ouchims@icrr.u-tokyo.ac.jp), Yuichi Harikane (UTokyo, hari@icrr.u-tokyo.ac.jp)

**Abstract:**
Pair-instability supernovae (PISNe) are explosions of very massive stars that may have played a critical role in the chemical evolution and reionization of the early Universe. In order to quantify their roles, it is required to know the PISN event rate at $z > 6$. Although Roman Space Telescope has a capability to discover PISNe at $z > 6$, identifying rare high-redshift PISN candidates among many other transients is challenging. In order to efficiently identify PISN candidates at $z > 6$, we propose to add sparse F213 observations reaching 26.5 mag (or deeper) every half year in the High Latitude Time Domain Survey. By adding the F213 information, PISNe at $z > 6$ can be efficiently identified in the color-magnitude diagram.

The Universe was mostly composed of hydrogen and helium after the Big Bang. It is believed that core-collapse supernovae (SNe) spread heavier elements synthesized in massive stars shortly after the Big Bang. Recent JWST observations of high-redshift galaxies revealed that some galaxies already had 6-20% of solar metallicity at $z \sim 10$ (Hsiao et al. 2023). Only a small fraction of galaxies are expected to have such a high metallicity at $z \sim 10$ (e.g., Ucci et al. 2023), and thus it is possible that the chemical enrichment in the early Universe proceeded much faster than that can be explained by canonical core-collapse SNe.

A possible missing link in our understanding of the chemical enrichment in the early Universe is pair-instability supernovae (PISNe). PISNe are theorized thermonuclear explosions of very massive stars with the zero-age main-sequence mass of around 150 - 300 Msun when there is no mass loss (e.g., Heger & Woosley 2002). Cosmological simulations of the first star formation suggest that first stars may be dominated by such massive stars exceeding 150 Msun (e.g., Hirano et al. 2015), and PISNe may have commonly existed in the early Universe. PISNe can produce more than 100 times more metals than typical core-collapse SNe. In addition, they can be more than 10 times more energetic than typical core-collapse SNe. Such energetic explosions may have also contributed to provide high-energy photons that led to the reionization of the early Universe. PISNe can make a profound difference in the early evolution of the Universe in many respects.

The fates of massive stars in the early Universe also provide a fundamental impact on black hole mass distributions observed by gravitational waves, because it takes a long time for binary black holes to merge. Coalescing binary black holes currently observed were formed in the early Universe. Because PISNe do not leave black holes behind, the existence of PISNe predicts that there are few black holes with masses between $\sim 50$ Msun and $\sim 120$ Msun (e.g., Farmer et al. 2020). Black hole mass distributions from gravitational wave observations reveal that there might be a significant decrease in the black hole mass distribution above $\sim 50$ Msun, but it is still not conclusive (LIGO-Virgo-KAGRA Collaboration 2021). PISN event rates at high redshifts can be compared to black hole merger rates and mass distributions from gravitational wave observations, so that we can discuss the evolutionary channel leading to close binary black holes.

In spite of the possible critical roles that PISNe may have played in the early Universe, we have not captured a smoking gun for their existence yet. Among thousands of SNe discovered by extensive modern transient surveys, only one promising PISN candidate SN 2018ibb has been finally discovered so far (Schulze et al. 2023). The rareness of PISNe in the current, local Universe is likely caused by high metallicity. Massive stars need to keep high mass throughout their life to explode as PISNe. Because mass-loss rates become higher at higher metallicities, it is likely that PISN event rates are much higher in the early Universe where metallicity is low (e.g., Briel et al. 2022). The upcoming Euclid mission has a potential to discover PISNe up to $z \sim 3.5$, and we may have many PISN discoveries in the coming years (Moriya et al. 2022a). Euclid will allow us to constrain PISN event rates up to $z \sim 3.5$ as well as their explosion properties. However, Euclid does not allow us to discover PISNe at $z > 3.5$.

In order to quantify the contributions of PISNe to the chemical evolution and reionization in the early Universe, it is essential to know their event rates at $z > 6$. Discovering even one PISN at $z >$

6 would be impactful to confirm the predicted existence and fate of the most massive stars at the epoch of reionization. Searching for SNe at such high redshifts requires wide-field, deep transient surveys in near-infrared. The Roman Space Telescope is an ideal instrument that provides an unprecedented opportunity to search for PISNe at z > 6.

The expected event rates of PISNe are much smaller than those of other typical SNe (e.g., Pan et al. 2012, Regős et al. 2020). While thousands of Type Ia SNe at z < 3 are expected to be discovered in the 5 deg$^2$ Deep Tier of the two-year High Latitude Time Domain Survey (Rose et al. 2022), only 1-10 PISNe at z > 6 are expected to appear (Moriya et al. 2022b). Therefore, it is essential to establish an efficient method to identify a couple PISN candidates among thousands of SNe.

There are several characteristics that can differentiate PISNe from other SNe. One is the long duration of PISNe. High-redshift PISNe can be bright for several years (Fig. 1), while other SNe are only bright for several months. In order to have a long light-curve coverage, it is desirable to have a long baseline deep time-domain data lasting for more than 2 years which are currently planned in the High Latitude Time Domain Survey, as proposed in another white paper by Fox et al. However, there are some other SNe with long durations such as Type IIn SNe. If available, host galaxy information would be helpful to identify high-redshift SNe. Still, many SNe can appear to be hostless, or can occur in a place where host galaxy identification is difficult.

In this white paper, we propose to add sparse F213 filter observations in the High Latitude Time Domain Survey to acquire additional, critical information that allows us to efficiently identify high-redshift PISN candidates. It has been shown that high-redshift PISNe can be distinguished from other SNe in color magnitude diagrams if we include the F213 photometry information (Fig. 2). The PISN candidate screening by color magnitude diagrams can be done even with a single epoch observation. In addition, high-redshift PISNe are slowly evolving. Thus, we do not require frequent observations in F213. For example, we can replace a F184 = 26.7 mag observation to a F213 = 26.5 mag (or deeper) observation every half year in the Deep Tier of the High Latitude Time Domain Survey. The effects of replacing F184 to F213 every half year on the cosmological parameter estimation with Type Ia SNe in the High Latitude Time Domain Survey needs to be investigated, but such a small change is not likely to affect the results significantly. Replacement in once a year would still be helpful.

The additional F213 observations can also be used to identify high-redshift superluminous SNe (SLSNe, Fig. 2). Although the origin of SLSNe is still not clear (e.g., Moriya et al. 2019), their event rates at high redshifts can provide a metallicity dependence of their occurrence rates, for example. We may discover some SLSNe at z > 6 (Moriya et al. 2022b). PISNe and SLSNe likely come from different mass ranges of massive stars and their relative fractions may provide some information on the initial mass functions in the early Universe.

The identified PISN and SLSN candidates through the proposed F213 observation can be followed by JWST and other ground-based facilities to confirm their redshifts and nature.

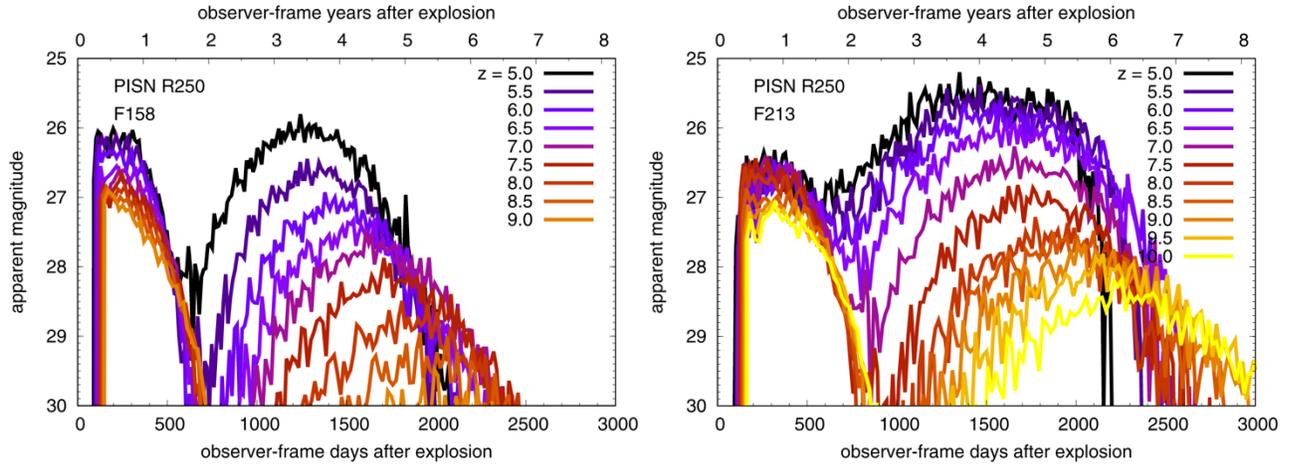

**Figure 1: Theoretical light curves of a PISN at high redshifts in the F158 (left) and F213 (right) filters (Moriya et al. 2022b). A PISN light curve model with the initial mass of 250 Msun that explodes as a red supergiant from Kasen et al. (2011) is adopted.**

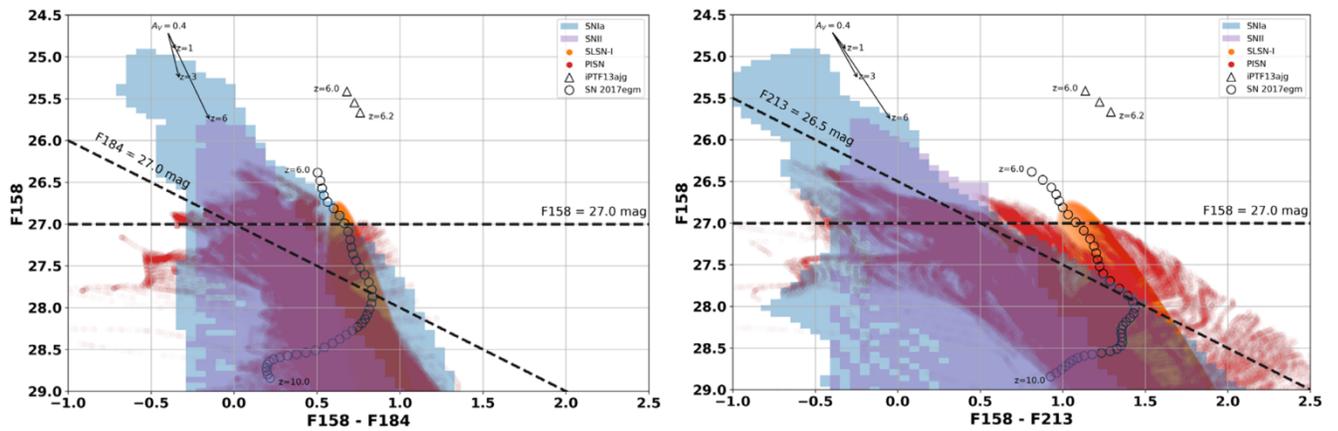

**Figure 2: Color-magnitude diagrams in F158-F184 (left) and F158-F213 (right). F213 is required to distinguish high-redshift PISNe and SLSNe from other low-redshift SNe (Moriya et al. 2022b).**